\documentclass[pra,twocolumn,showpacs,twoside,superscriptaddress]{revtex4}
\usepackage{graphicx}
\usepackage{amsmath,amsthm}
\usepackage{placeins}
\usepackage{color}
\usepackage{hyperref}
\usepackage{bbm}
\newcommand{\eq}{\begin{eqnarray}}
\newcommand{\en}{\end{eqnarray}}

\def\bra#1{\mathinner{\langle{#1}|}}
\def\ket#1{\mathinner{|{#1}\rangle}}
\newcommand{\braket}[2]{\langle #1|#2\rangle}

\begin{document}

\title{Bosonic behavior of entangled fermions } 

\author{Malte C. Tichy}
\affiliation{Lundbeck Foundation Theoretical Center for Quantum System Research,
Department of Physics and Astronomy, University of Aarhus, DK-8000 Aarhus C, Denmark}

\author{Peter Alexander Bouvrie}
\affiliation{Departamento de F\'isica At\'omica, Molecular y Nuclear and Instituto Carlos I de F\'isica Te\'orica y Computacional, Universidad de Granada, E-18071 Granada, Spain}

\author{Klaus M\o{}lmer}
\affiliation{Lundbeck Foundation Theoretical Center for Quantum System Research,
Department of Physics and Astronomy, University of Aarhus, DK-8000 Aarhus C, Denmark}

\date{\today}


\begin{abstract}
Two bound, entangled fermions form a composite boson, which can be treated as an elementary boson as long as the Pauli principle does not affect the behavior of many such composite bosons. The departure of ideal bosonic behavior is quantified by the normalization ratio of  multi-composite-boson states. We derive the two-fermion-states that extremize the normalization ratio for a fixed single-fermion purity $P$, and establish general tight bounds for this indicator. For very small purities, $P< 1/N^2$, the upper and lower bounds converge, which allows to quantify accurately the departure from perfectly bosonic behavior, for any state of many composite bosons. 
\end{abstract}

\maketitle

\section{Introduction}
The composition principle, the treatment of composite particles as elementary objects despite their underlying structure, is a fundamental pillar of natural science  \cite{Healey2012}. In the microscopic world governed by quantum mechanics, from hadrons at the highest achievable energies \cite{Khachatryan2010} to ultracold molecules \cite{Zwierlein2003}, the composition principle allows us to treat particles  with integer spin as elementary bosons. This treatment can greatly simplify the understanding of the many-body behavior and the statistical physics of composite bosons; but when can we confidently apply it and reliably treat two bound fermions as a boson?

The hierarchy of energy scales in nature would appear to provide an answer, and one intuitively expects that the bosonic behavior of bound fermions  relies on their strong binding. For example, weakening the bound between fermions indeed leads to the BEC-BCS crossover \cite{Bourdel2004}. However, even when the constituents of \emph{cobosons} \cite{Combescot2007}, {\it i.e.}~of compounds constituted of two fermions, are perfectly bound, it is not granted that the creation and annihilation operators of cobosons obey the bosonic commutation relations and exhibit perfect bosonic behavior: The Pauli principle for the underlying constituents may become relevant and thus jeopardize bosonic dynamics  \cite{Law2005,Chudzicki2010,avancini,Rombouts2002,Combescot2007,Combescot2008a,Combescot2010,Combescot2011a}. For good bosonic behavior, the occupation probability of any single-fermion state must be low, such that the constituent fermions of the cobosons do not compete for available single-fermion states. 

A satisfactory answer to the above question was given using the tools of quantum information \cite{Law2005,Chudzicki2010}: Independently of the actual physical system and of the binding strength between the constituents, the bosonic behavior of cobosons is intimately related to the \emph{entanglement} between the constituting fermions \cite{Law2005}, and the impact of the Pauli principle fades away with increasing entanglement \cite{Law2005,Sancho2006,Chudzicki2010,Ramanathan2011,Gavrilik2012}.  As an indicator for the entanglement between the fermions, one may use the purity $P$ of the reduced states of either fermion \cite{Tichy2011a}: $1/P$ is the effective number of Schmidt modes, {\it i.e.}~of populated single-fermion states. To treat cobosons as ideal bosons and evade the Pauli principle, there need to be many more single-particle states ($1/P$) than composites ($N$), {\it i.e.}~$N \cdot P \ll 1$ \cite{Law2005}. The original argument in \cite{Law2005} was based on specific wavefunctions, it was generalized to arbitrary states in \cite{Chudzicki2010}, where general upper and lower bounds to the indicator of bosonic behavior -- the ratio of normalization constants of many-coboson states that will be introduced below -- were found. 

Here, we strengthen further the relationship between entanglement, as characterized by the purity of the single-fermion states, and the compositeness character of cobosons: We improve on existing bounds and derive the explicit form of those quantum states that maximize and minimize the normalization ratio for a given purity $P$. Our bounds are optimal, since they are saturated by the extremal states found. The tight upper bound comes close to the lower bound when $ N^2\cdot P \ll 1$, {\it i.e.}~in this regime, not only is the deviation from perfectly bosonic behavior small, but it can also be bound very tightly via the purity. 

We first present the formalism for the treatment of $N$-coboson states and review previous results on the normalization ratio, in Section \ref{manycobostate}, following the notation of \cite{Law2005,Chudzicki2010}. Our main result, tight bounds for the normalization ratio, is derived in Section \ref{tightbounds}. Examples and a discussion of limiting extremal cases are then given in Section \ref{interpret}. We conclude with a combinatorial interpretation of the findings and an outlook on possible extensions and applications in Section \ref{concluoutl}.

\section{Many-coboson states} \label{manycobostate}

\subsection{Normalization of many-coboson states and entanglement}
We follow the formalism of \cite{Law2005,Chudzicki2010} for cobosons that are constituted of two fermions. The creation operator for a coboson constituted of distinguishable fermions can always be written in the Schmidt decomposition \cite{Law2005,Chudzicki2010}
\eq 
\hat c^\dagger = \sum_{j=1}^S \sqrt{\lambda_{j}} ~ \hat a_{j}^\dagger \hat b_{j}^\dagger 
 , \label{compositeboson}
\en
{\it i.e.}~as a sum over only one index, where the $\lambda_j$ are the Schmidt coefficients, $\hat a^\dagger_{j}$ ($\hat b^\dagger_{j}$) creates an $a$ ($b$) -type fermion in the Schmidt mode $j$, and the number of Schmidt coefficients is denoted by $S$. For cobosons composed of two indistinguishable fermions, we use the Slater decomposition \cite{SlaterDecomposition} 
\eq 
\hat c^\dagger = \sum_{j=1}^{2S} \sqrt{\lambda_{j}} ~ \hat f_{2j-1}^\dagger \hat f_{2j}^\dagger .
\en
and set $\hat a_{j} := \hat f^\dagger_{2j-1}$ and $\hat b_j := \hat f^{\dagger}_{2j}$ to recover the form (\ref{compositeboson}). 

A state of $N$ composite bosons reads
\eq 
\ket{N}=\left( \chi_N \right)^{-1/2} \frac{\left( \hat c^\dagger \right)^N }{\sqrt{N!}} \ket{0} .
\en
where the coboson normalization factor $\chi_N$ witnesses the possible departure from the familiar bosonic behavior \cite{Combescot2003,Law2005,Chudzicki2010,Combescot2010,Combescot2011}.  The factor $\chi_N$, leading to the normalization of the many-coboson state, $1=\braket{N}{N}$, is a completely symmetric polynomial in the Schmidt coefficients $\lambda_j$ \cite{Macmahon1915},
\eq 
\chi_N 
= N! \sum_{1\le j_1< \dots < j_N \le S }~ \prod_{m=1}^N \lambda_{j_m} . \label{chidef}
\en
 For ideal bosons, $\chi_N=1$ for all $N$, while $\chi_N=0$ when the number of cobosons $N$ is larger than the number of available fermionic single-particle states, $S$.

The probability distribution $\vec \lambda$ is characterized by its power-sums   \cite{Macmahon1915,Ramanathan2011,Combescot2003}
\eq M(m)=\sum_{j=1}^S \lambda_j^m ,  \label{powersum} \en 
where $M(1)=1$ due to normalization, $0<M(2)=P\le 1$ is the purity of the distribution $\vec \lambda$. Power-sums are also called frequency moments, they are directly related to the R\'enyi entropy of the distribution $\vec \lambda$, $ H_{\text{R\'enyi}}^m(\vec \lambda)=   \text{log} \left(M(m) \right)/(1-m) .$ 
The $M(m)$ are independent, but Jensen's and H\"older's inequalities \cite{Hardy1988} apply: 
 \eq 
 M(k-1)^{\frac{k-1}{k-2}}  \le M(k) \le   M(k-1)^{ \frac k {k-1}} . \label{powersumconstraint}
\en
Using power-sums, the normalization constant  can be expressed recursively \cite{Ramanathan2011}
\eq
\chi_N={(N-1)!} \sum_{m=1}^N \frac{(-1)^{1+m}}{(N-m)!} M(m) \cdot \chi_{N-m}, \label{chinpwsums}
\en
where we set $\chi_0=1$ for convenience.

Combinatorially speaking, the quantity $\chi_N$ is the probability that, given a set of $N$ objects which are each randomly assigned a property $j$ (with $1\le j \le S$) with probability $\lambda_j$, all objects carry different properties. For example, for $S=365$ and $\lambda_j=1/365$, we obtain the solution to the ``birthday problem'', {\it i.e.}~the likelihood that all members of a group of $N$ people have a different birthday. The power-sum $M(m)$ is the probability that, selecting $m$ objects that each carry a property distributed according to $\lambda_j$, all $m$ objects have the same property. Therefore, we have the simple relationship $\chi_2=1-M(2)$, while the $\chi_N$ with $N\ge 3$ are functions of all $M(m)$ with $m\le N$, as given by (\ref{chinpwsums}).

\subsection{Normalization ratio as a measure of  bosonic behavior}
The normalization factor  $\chi_N$ of an $N$-coboson state reflects the probability to create a state of $N$ cobosons by the $N$-fold application of the coboson creation operator on the vacuum.  The resulting state reads 
\eq 
\frac{  \left(\hat c^\dagger \right)^N }{\sqrt{N!}}\ket 0 =\sum_{j_1\neq j_2 \dots \neq j_N}^{1 \le j_m \le S} \left(  \prod_{k=1}^N \sqrt{\lambda_{j_k}} \hat a_{j_k}^\dagger \hat b_{j_k}^\dagger  \right) \ket{0} ,
\en
{\it i.e.}~it is a weighted superposition of \emph{all} possibilities to distribute the pairs of fermions among the pairs of Schmidt modes (for species $a$ and $b$). Every pair of Schmidt modes $a_{j}^\dagger \hat b_{j}^\dagger \ket{0} $ is -- at most -- occupied by one fermion pair, in close analogy to the birthday problem. 

The normalization \emph{ratio}, $ \chi_{N+1}/\chi_N$, has emerged as a decisive indicator for the bosonic behavior of a state of $N$ cobosons: For example, adding an additional coboson to an $N$-coboson state, {\it i.e.}~applying the coboson creation operator, leads to the state \cite{Chudzicki2010}
\eq 
\hat c^\dagger \ket{N} = \sqrt{\frac{\chi_{N+1}}{\chi_N} } \sqrt{N+1} \ket{N+1} ,
\en
{\it i.e.}~the state is sub-normalized: It is possible that the addition of the $(N+1)$st coboson is inhibited by the Pauli principle, which occurs with probability ${1-\chi_{N+1}/\chi_N}$.  Similarly, the departure of the expectation value of the commutator $[ \hat c, \hat c^\dagger ]$, which is unity in the ideal case, reads \cite{Law2005}
\eq 
\bra N [ \hat c, \hat c^\dagger ] \ket N = 2 \frac{\chi_{N+1}}{\chi_N} -1  .
\en
The evaluation of $\chi_N$ scales prohibitively in the number of particles $N$, even when using the recursive formula (\ref{chinpwsums}) \cite{Klotz1979}. Approximations to the normalization factor in terms of easily accessible quantities, such as the purity $P\equiv M(2)$, are thus desirable. From Eq.~(\ref{chinpwsums}) and for small  $N\cdot P$, a series expansion can be derived \cite{Ramanathan2011,Combescot2011a},
\eq 
\frac{\chi_{N+1}}{\chi_N} &\approx & 1 - N\cdot P + N^2(M(3)-P^2) \nonumber \\ && + \mathcal{O}\left(N^3(M(4)+2 P^3-2 P~ M(3) )\right) . \label{seriesexp}
\en 
On the other hand, an upper and a lower bound to the normalization ratio were found \cite{Chudzicki2010}, 
\eq 
1- P\cdot N \le \frac{\chi_{N+1}}{\chi_N} \le 1-P . \label{chudine}
\en
However, the upper bound $1-P$ is independent of $N$ and cannot be saturated, and the form of typical states that maximize the ratio is not known. Here, we derive tight bounds, find the quantum states that saturate these bounds, and give a physical interpretation for their optimality.

\section{Tight bounds on  the normalization ratio} \label{tightbounds}

\subsection{Extremal entangled states}

We are interested in the possible values of the normalization ratio $\chi_{N+1}/\chi_{N}$ for a given $P$. In order to find the extremal values of $\chi_{N+1}/\chi_{N}$,  we maximize and minimize this quantity under the constraints $M(1)=1, M(2)=P$. 

Given a finite $P$, the number $S$ of non-vanishing $\lambda_j$ is bound from below by $L$, the smallest integer that is equal to or larger than $1/P$:  \eq S \ge L := \left \lceil \frac 1 P \right \rceil. \en

Distributions $\vec \lambda$ with $S-1$ equal coefficients \cite{Wei2003} constitute extremal states, and will turn out to minimize or maximize $\chi_{N+1}/\chi_N$.  We thus define
\eq 
\lambda_{1}^{(\pm)} &=& \frac{ 1\pm\sqrt{ (S-1) (S P-1)} } S , \nonumber  \\  
\lambda_{j \in \{ 2\dots S\}}^{(\pm)} &=& \frac{1-\lambda_1^{(\pm)} }{S-1} . \label{uniformandpeaked}
\en
When we choose $S=L$, we obtain the \emph{uniform distribution} $\vec \lambda^{(u)}:=\vec \lambda^{(-)}_{(S=L)}$ with $\lambda_1^{(u)}\le \lambda_{j \in \{ 2,\dots,L\}}^{(u)}$. This distribution minimizes the number of non-vanishing Schmidt coefficients. 

\begin{figure}[ht]
\center
\includegraphics[width=\linewidth,angle=0]{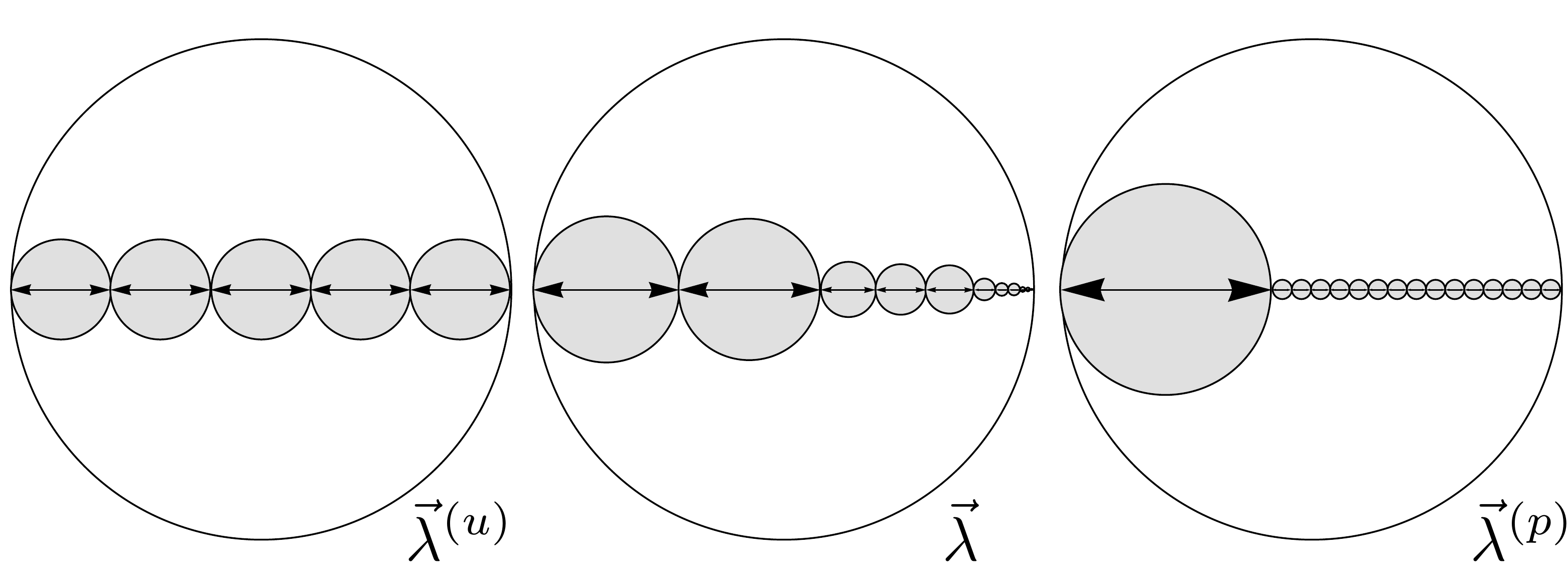} 
\caption{Visualization of probability distributions $\vec \lambda$ with a maximal number of Schmidt coefficients ${S \le 16}$ and purity $P=1/5$. The diameter of each filled circle represents the probability $\lambda_j$, its area is proportional to $\lambda_j^2$. The diameters sum to unity, while the total area adds up to $P$. Distributions with the same occupied gray area thus have the same purity $P$,  the three distributions shown cannot be distinguished via their purity $P$, but only through higher-order power-sums. Given $P=1/5$, different distributions $\vec \lambda$ can lead to different normalization factors $\chi_N$. The uniform distribution $\vec \lambda^{(u)}$ minimizes the normalization ratio, while the peaked distribution $\vec \lambda^{(p)}$  maximizes it under the chosen constraint $S\le 16$. } 
 \label{VisuDistribu}
\end{figure}

For $S\ge L$, we find a \emph{peaked distribution} $\vec \lambda^{(p)}:=\vec \lambda^{(+)}_{(S\ge L)}$ that satisfies $\lambda_1^{(p)}\ge \lambda_{j \in \{ 2,\dots,S\}}^{(p)}$. In the limit $S \rightarrow \infty$, the peaked coefficient $\lambda^{(p)}_1$ converges to $\sqrt{P}$, while all other coefficients become vanishingly small, while the distribution always remains normalized and possesses the purity $P$. Choosing $P=1/S$ implies $L=S$ and $\vec \lambda^{(p)}=\vec \lambda^{(u)}$ -- all coefficients are then identical, a maximally entangled state is obtained. In Fig.~\ref{VisuDistribu}, we illustrate the uniform distribution $\vec \lambda^{(u)}$, a randomly chosen distribution $\vec \lambda$, and the peaked distribution $\vec \lambda^{(p)}$ with the same purity $P=1/5$ and $S \le 16$.

The peaked and the uniform distributions have extremal properties: For example, they saturate the bounds on the higher-order power-sums $M(k \ge 3)$, given by (\ref{powersumconstraint}). In the limit $S \rightarrow \infty$, we have 
\eq 
M^{(p)}(k) = \sqrt P\cdot M^{(p)}(k-1)  ,
\en
and, for fractional values $P=1/L$,  we find for the uniform distribution:
\eq 
M^{(u)}(k) =  P\cdot M^{(u)}(k-1) .
\en
Combinatorially speaking, given the probability to find a pair of objects with the same property ({\it i.e.}~given the value $P=M(2)$), the probability $M(3)$ to find three  objects with the same property for three randomly chosen objects is maximized by the $\vec \lambda^{(p)}$ and minimized by the $\vec \lambda^{(u)}$ distributions.

\subsection{Normalization ratio for extremal states}
Since only two different non-vanishing values of $\lambda_j$ appear for the uniform and the peaked distributions, the normalization ratio can be computed explicitly for these extremal distributions:
\begin{widetext}
\eq 
\frac{\chi^{(p)}_{N+1}}{\chi^{(p)}_N}& = &\frac{(S-N) (1-P) \left(S-1 + N \sqrt{(S-1) (S P-1)}\right)}
{(S-1) \left(S+S (N-1) P-N \left(1 - \sqrt{(S-1) (S P-1)}\right)\right)} , \label{ratioextrdistp} \\
\frac{\chi^{(u)}_{N+1}}{\chi^{(u)}_N} & =& 
\frac{(L-N) (1-P) \left(L-1 - N \sqrt{(L-1) (L P-1)}\right)}
{(L-1) \left(L+L (N-1) P-N \left(1 + \sqrt{(L-1) (L P-1)}\right)\right)} 
, \label{ratioextrdistu}
\en
where $\chi^{(u/p)}_N$ is the normalization factor for the uniform/peaked distribution. We can now formulate our tight bounds on $\chi_{N+1}/\chi_N$, given an arbitrary distribution $\vec \lambda$ of $S$ Schmidt coefficients: 
\eq 
1-P\cdot  N 
 \stackrel{(i)}{\le} \frac{\chi^{(u)}_{N+1} } { \chi^{(u)}_{N} }
  \stackrel{(ii)}{\le} \frac{\chi_{N+1} } { \chi_{N} }
    \stackrel{(iii)}{\le} \frac{\chi^{(p)}_{N+1} } { \chi^{(p)}_{N} } 
      \stackrel{}{\le}  \lim_{S\to \infty}  \frac{\chi^{(p)}_{N+1} } { \chi^{(p)}_{N} }  
\stackrel{(iv)}{=} \mathcal{U}_N(P) 
       \stackrel{(v)}{\le} 1-P \label{biginequality} ,
\en
\end{widetext}
where $\chi^{(p)}_{N+1}/\chi^{(p)}_{N}$ is computed for the finite $S$ defined by $\vec \lambda$, and we define the upper bound 
\eq \mathcal{U}_N(P) = 1 -\frac{P\cdot  N}{1 + (N-1) \sqrt{P} }  \label{upperbound}  .\en
The inequalities (\ref{chudine}) are represented here by the extremal lower and upper bounds $(i,v)$, they were first shown in \cite{Chudzicki2010}, for an alternative proof see \cite{Combescot2011}. We  prove the new bounds $(ii - iv)$ in Appendix \ref{proofs}, and discuss their physical implications in the following Section \ref{interpret}.

\section{Illustration and interpretation of bounds} \label{interpret}
\begin{figure}[ht]
\center
\includegraphics[width=\linewidth,angle=0]{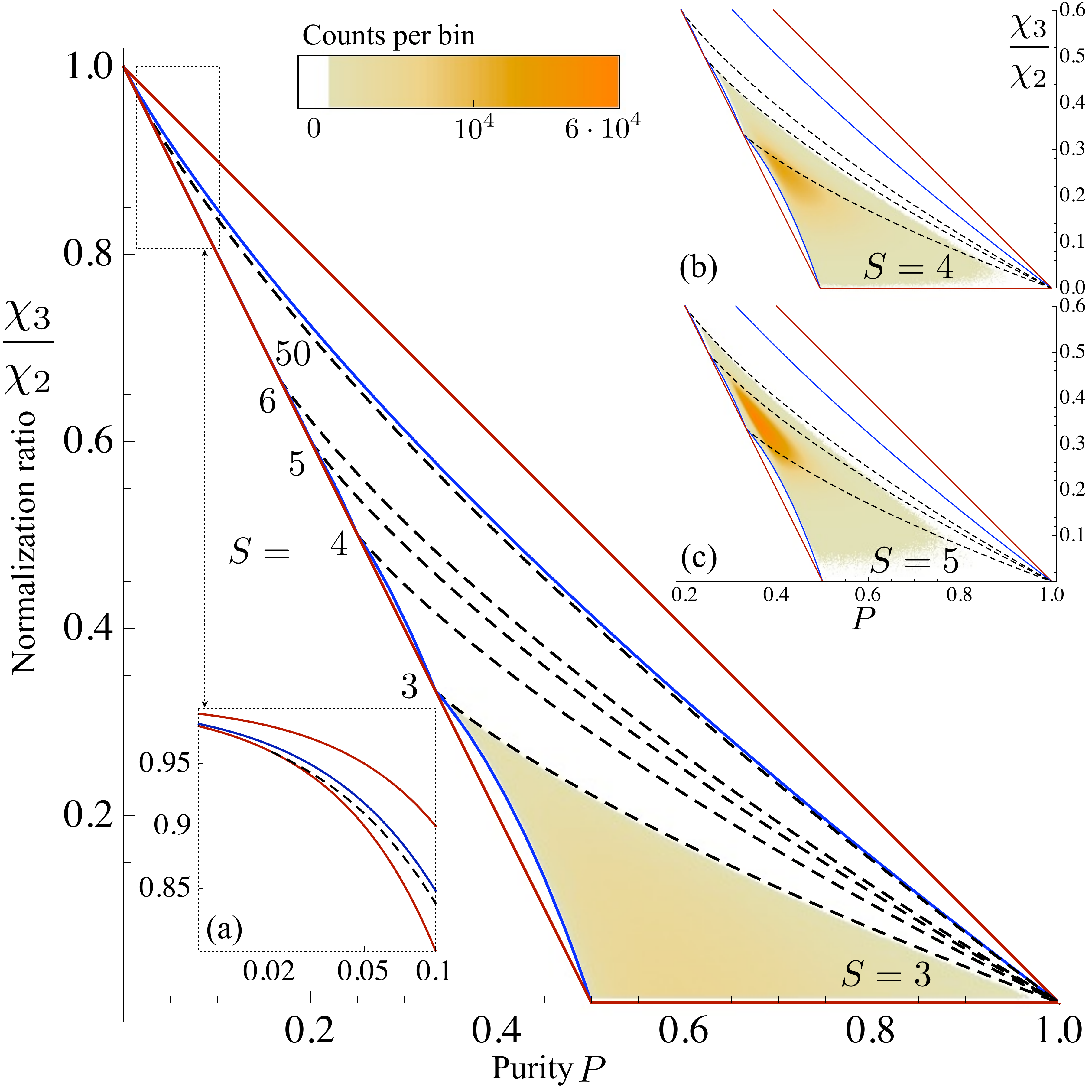} 
\caption{(color online) Bounds for the normalization ratio $\chi_3/\chi_2$ as a function of the purity $P$. The red solid lines indicate the extremal bounds in Eq.~(\ref{biginequality}) found in \cite{Chudzicki2010}. The solid blue lines denote the tight bounds in Eq.~(\ref{biginequality}) that can be achieved for any value of $P$. For a given maximal number of Schmidt-coefficients $S$, the black dashed lines are the corresponding upper bounds (for peaked states $\lambda^{(p)}$), which merge with the lower bound at $P=1/S$ . Inset (a): The scale in $P$ is logarithmic, the (black dashed) upper bound for $S=50$ is close to the total upper bound at $P=1/10$, but it merges with the lower bound at $P=1/50$. The distributions show the numerically obtained values for the normalization factor and the purity, for different fixed Schmidt numbers, $S=3$ (main figure) and $S=4,5$ (insets (b), (c)). } 
 \label{BoundsIllu}
\end{figure}

All bounds for $N=2$ are illustrated in Fig.~\ref{BoundsIllu}.

\subsection{Upper and lower bounds}
The authors of \cite{Chudzicki2010} showed that the lower bound ${1-N\cdot P}$ in (\ref{chudine}) is attained for fractional values of $P$, {\it i.e.}~for $P=1/L=1/S$. Setting $L=1/P$ in (\ref{ratioextrdistu}) reproduces the bound, and ${\chi_{N+1}^{(u)}}/{\chi_N^{(u)}}=1-N\cdot P$. The saturation can also be observed in Fig.~\ref{BoundsIllu}: The tight lower bound (blue line) coincides with $1-N\cdot P$ (red line) for fractional values of the purity $P=1/S$. When $P\ll 1$, and thus $P\approx 1/L$, the tight bound (\ref{ratioextrdistu}) differs only marginally, while for large purities $P \lesssim 1/N$ the tight bound can differ significantly from $1- N\cdot P$. In the limit $P \rightarrow 1/N$, the tight bound (\ref{ratioextrdistu}) and the previous bound $1-N\cdot P$ differ by a factor $(1+N)/2$.  

In contrast to the previously established upper bound $1-P$, our tight upper bound (\ref{biginequality},$iv$) depends on the number of particles $N$. When the number of non-vanishing Schmidt coefficients $S$ is finite, the bound (\ref{ratioextrdistp}) is more efficient than the limiting case $S \rightarrow \infty$: In Fig.~\ref{BoundsIllu}, the dashed black lines show the upper bound for finite $S$, while the blue line indicates the absolute upper bound. When the purity is decreased for a constant $S$, the upper and lower bounds eventually merge when $P=1/S$ is attained (see also inset (a)).

\begin{figure}[ht]
\center
\includegraphics[width=\linewidth,angle=0]{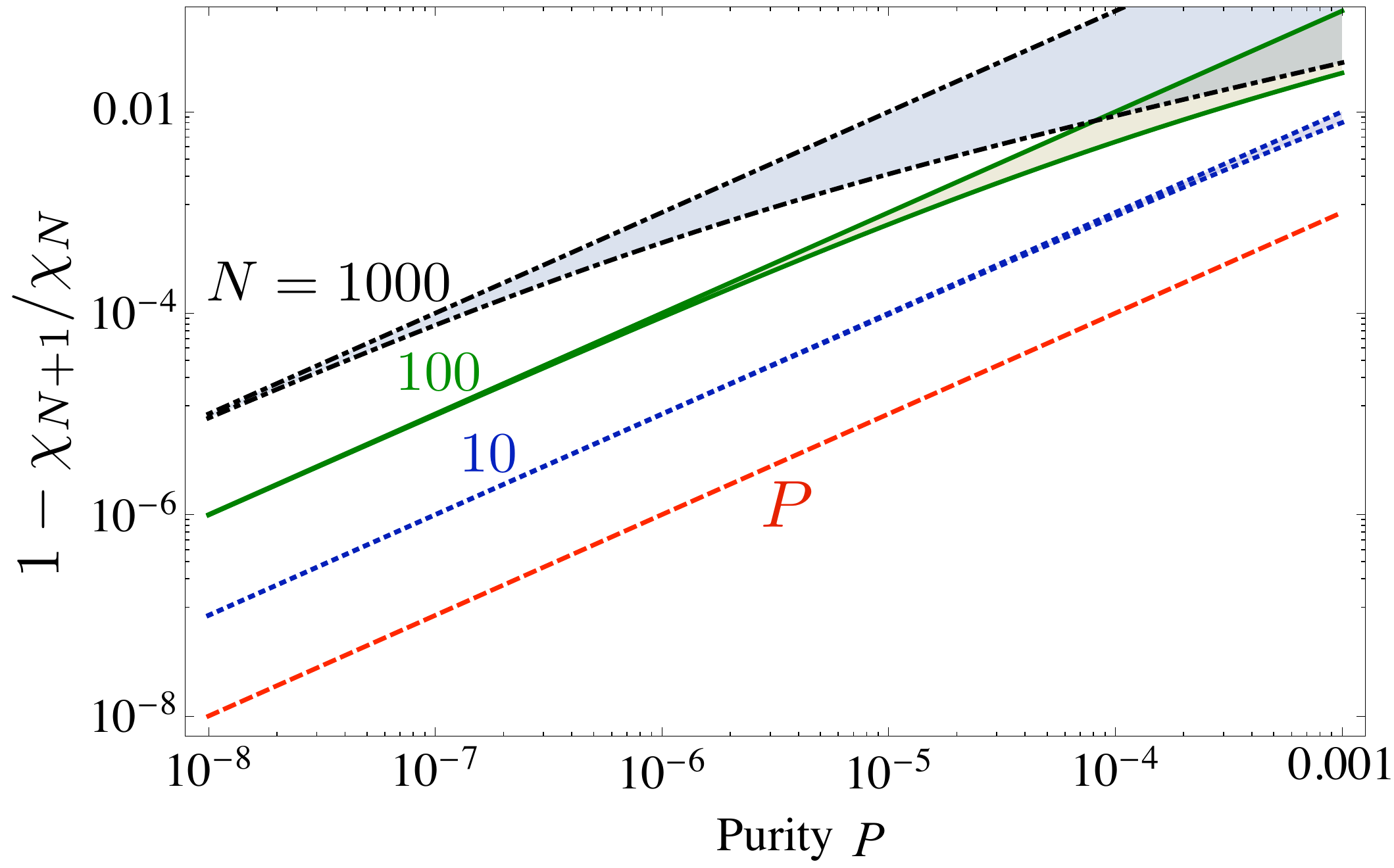} 
\caption{(color online) Deviation of the normalization ratio from unity, $1-\chi_{N+1}/\chi_N$, as a function of the purity $P$, in log-log-representation. The red dashed line is the $N$-independent upper bound to $\chi_{N+1}/\chi_N$, {\it i.e.}~here it represents the lower bound $P$ to the deviation. The black dot-dashed, green solid and blue dotted lines indicate the tight bounds for $N=1000$, $N=100$, and $N=10$, respectively. The previously found  bound $1-N\cdot P$ \cite{Chudzicki2010} and the tight bound $(\ref{ratioextrdistu})$ do not differ significantly in the present regime $P \ll 1$, they thus cannot be distinguished in the plot. The normalization ratio can take any value in the respective shaded areas. The deviation from ideal behavior as well as the gap between the upper and lower bound decrease with decreasing $P$. } 
 \label{OnemChiplot}
\end{figure}

The upper bound can be expanded in powers of $\sqrt{P}$:
\eq 
\mathcal{U}_N(P) & =&  1 -  \sum_{k=2}^\infty  P^{k/2} \frac{(1-N)^k N}{(N-1)^2} \\ &=& 1 -\lim_{n\rightarrow \infty}  \frac{N ( (N-1) P+ (1 - N)^n P^{(n+1)/2})}{( N-1) (1 +(N-1)  \sqrt{P} )}  , \nonumber    
\en
with convergence radius $P < 1/(N-1)^2$. To second order, we then find 
\eq 
\mathcal{U}_N(P) \approx 1-P\cdot N + P^{3/2} (N^2-N) + \mathcal{O}(P^{2}), \en {\it i.e.}~the upper and lower bounds coincide in the limit $P\rightarrow 0$. Indeed, for $P \ll 1/(N-1)^2 \approx 1/N^2$, the denominator in (\ref{upperbound}),  $1+(N-1)\sqrt{P}$, is of the order of unity. This behavior is illustrated in Fig.~\ref{OnemChiplot}, where we plot the deviation from perfect bosonic behavior, ${1-\chi_{N+1}/\chi_N}$. The $N$-dependence of the new upper bound is apparent, as well as the convergence of the upper bound to the lower bound. In particular, the purity essentially defines the normalization ratio in the range $P\cdot N^2 \ll 1$.

In order to illustrate the typical behavior of random cobosons, we generated $3\cdot 10^8$ random distributions $\vec \lambda$ \cite{Giraud2007}, sampled according to the Haar-measure \cite{Zyczkowski2011,Giraud2007,Page1993}. Pairs ($P, \chi_{3}/\chi_2$) are counted in a grid with 1000$\times$1000 bins, which is translated to a color-code in Fig.~\ref{BoundsIllu}. We generated states with $S=3$ (main figure), $S=4$ (inset (b)) and $S=5$ (inset (c))  non-vanishing Schmidt coefficients. The bounds for finite $S$ are indeed reached by randomly generated states. We also observe a concentration of states around the peak value of $(P, \chi_3/\chi_2)$ when the number of Schmidt modes $S$ is increased: The vast majority of randomly generated states in high dimensions share very similar entanglement properties \cite{Hayden2006}.

\subsection{Limit of many particles}
Surprisingly, increasing the number of particles at constant purity $P$ does not always fully destroy bosonic behavior: The lower bound $1-N\cdot P$ admittedly decreases with $N$ when $P>0$ and vanishes for $P=1/N$ -- the corresponding uniform state $\lambda^{(u)}$ consists of a finite number $L=\left \lceil 1/P \right \rceil$ of Schmidt modes, such that at most $L$ particles can be accommodated and $\chi_{L+1}=0$. The peaked state, however, leads to non-vanishing $\chi_{N+1}/\chi_{N}$   for arbitrarily large particle numbers: In the limit $N \rightarrow \infty$, we have 
\eq 
 \mathcal{U}_N(P)  \stackrel{N \rightarrow \infty}{\rightarrow} 1- \sqrt{P} . \label{manyN}
\en
That is to say, for the peaked state the departure from bosonic behavior, as quantified by the ratio $\chi_{N+1}/\chi_N$, amounts to at most $\sqrt{P}$, for \emph{any} number of particles $N$. This counter-intuitive result can be understood by the extremal form of the distribution $\vec \lambda^{(p)}$: When a fermion pair populates a Schmidt mode other than the first one (which is populated with probability $\sqrt{P}$), it can essentially be neglected for the impact of the Pauli principle on the next fermion pair, since there are arbitrarily many such modes that are occupied with vanishing probability (in the limit $S\rightarrow \infty$). Assuming that $N \gg 1$ particles were successfully prepared, the probability that the first pair of Schmidt modes is populated by some fermion pair is $1-\sqrt{P}^N$, {\it i.e.}~very close to unity. Adding an $(N+1)$st particle is thus successful when this last particle does not end in the first Schmidt mode, {\it i.e.}~the success probability is $1-\sqrt{P}$, just as given in (\ref{manyN}). Colloquially speaking, there are always enough Schmidt modes to accommodate another particle. The last added particle must not, however, end in the first Schmidt mode, since the latter is occupied with nearly unit probability. 

Although an $N$-coboson state still behaves bosonic to a certain degree, the normalization factor for such a state, given the peaked distribution $\vec \lambda^{(p)}$, is 
\eq \chi_N^{(p)}=\left(1 - \sqrt{P} \right)^{N-1} \left(1 +(N-1)\sqrt{P} \right) , \en 
which converges to zero for any $0<P \le 1$ in the limit $N \rightarrow \infty$. The analogous normalization factor for a uniform distribution (assuming $P=1/L$) reads 
\eq
\chi_N^{(u)}=\frac 1  {L^N} \frac{L!}{(L-N)! } ,\en 
which decays faster in $N$ than $\chi_N^{(p)}$ and vanishes identically for $N>L$.

\section{Conclusions and Outlook} \label{concluoutl}
The $N$-coboson normalization ratio $\chi_{N+1}/\chi_N$ was established as an important indicator for the composite behavior of non-elementary bosons \cite{Law2005}. Our main result is a new, $N$-dependent upper bound for $\chi_{N+1}/\chi_N$, and the explicit representation of the states that saturate this bound. 

In the limit of small purities ${P<1/N^2}$, the bosonic behavior of an $N$-coboson state is very tightly defined by $P$, since the lower and upper bounds merge. In practice, the purity of a bound pair of particles often satisfies $P\ll 1/N^2$, {\it e.g.}~for atoms in a trapped BEC \cite{Rombouts2002,Chudzicki2010}. Our bounds thus provide a simple and reliable way to quickly check the departure of bosonic behavior of any type of cobosons. With a combinatorial argument, we can understand this clear determination of bosonic behavior: For small purities $P N^2 <1$, the probability to not finding $N$ objects with different properties is essentially determined by the probability to find exactly one pair of objects with the same property, which is defined by $M(2)$. Triplets and larger combinations that depend on higher-order $M(k)$ are then essentially negligible.

When the purity is not very small, $P\approx 1/N$, however, the form of the wavefunction does play a role for bosonic behavior, which is then not entirely defined by $P$. It might be possible to access entanglement properties of bound fermions via the higher power-sums $M(m)$, which  can be obtained by measuring $\chi_N$ \cite{Ramanathan2011a}. 

Formally speaking, we found bounds to the completely symmetric polynomial (\ref{chidef}) in terms of the first and second power-sums (\ref{powersum}), $M(1)=1$ and $M(2)\equiv P$. For tighter bounds, one could specify also the third power-sum, $M(3)$, and repeat the maximizing- and minimizing procedure of Section \ref{proofs} to find those states that minimize/maximize $\chi_{N+1}/\chi_{N}$ for given $M(1), M(2), M(3)$; this procedure could  be extended to even higher orders. It is, however, not immediate how operations that are analogous to (\ref{ope1}) and (\ref{ope2}) but which leave $M(3)$ invariant can be constructed. In the typically encountered domain of small purities, $P\ll 1/N^2$, this endeavor is not an urgent desideratum, since the encountered bounds as a function of $P$ are already tight. On the other hand, using relations between R\'enyi entropies of different orders \cite{IEE,karol}, our bounds can be re-formulated in terms of other indicators for entanglement, such as the Shannon entropy of $\vec \lambda$.

Given a fixed purity $P$, the uniform distribution $\vec \lambda^{(u)}$ minimizes the probability that the Pauli principle is irrelevant, while the peaked distribution maximizes it. In other words, the $N^2$-coefficient $(M(3)-P^2)$ in the expansion (\ref{seriesexp}) is maximized. Although a peaked or canyon distribution leads, in general, to a normalization ratio that is smaller than for the uniform distribution \cite{Combescot2011}, this is mainly due to the consequent change of purity. For fixed purity $P$, the bosonic behavior of the uniform distribution is actually inferior with respect to the peaked one. 

Combinatorially speaking, we have considered a variant of the birthday problem with non-uniform probabilities \cite{Klotz1979}. Here, Schmidt modes or single-particle quantum states take the role of birthdays \cite{Klotz1979,Munford1977} or surnames \cite{Mase1992}. Rather counter-intuitively, the optimal bosonic behavior for a fixed purity $P\equiv M(2)$ is found by \emph{maximizing} the probability to find three objects with the same properties, {\it i.e.}~$M(3)$. This result can be understood as follows: Any pair of objects that have the same property is as deleterious as any triplet (or any other $m$-tuplet). The probability to find a pair, however, \emph{decreases} with increasing $M(3)$. This decrease has a larger impact on the overall probability to find all objects with different properties than the consequent increase of the probability to find triplets with $M(3)$. For example, for $N=3$, the probability to find a pair amounts to $3(P-M(3))$, the probability to find a triplet is $M(3)$. The overall probability to find all objects with different properties amounts to $1-3 P + 2 M(3)$. 

To complement the analytical bounds, we have numerically generated random states, which do not only show a concentration around the most probable value of the purity $P$ \cite{Hayden2006}, but they also cluster around a certain value of the normalization ratio, consistent with the concentration-of-measure phenomenon. It remains to be studied how random states in higher dimensions behave in general, {\it i.e.}~what is the \emph{typical} normalization ratio and the distribution of states with a given purity. 

\subsection*{Acknowledgements}
This work was partially supported by the Project FQM-2445 of the Junta de Andaluc\'ia and the grant FIS2011-24540 of the Ministerio de Innovaci\'on y Ciencia, Spain.
\appendix

\section{Proof of bounds  and tightness}\label{proofs}


In order to show Eq.~(\ref{biginequality}) $(ii)$ and $(iii)$, we construct operations on the distribution $\vec \lambda$, in Section \ref{operations}. These leave the sum of the $\lambda_j$ and the purity $P$ invariant, while they increase or decrease the normalization ratio, as shown in Section \ref{normrationunder}. Since only the extremal distributions $\vec \lambda^{(p)}$ and  $\vec \lambda^{(u)}$ remain invariant under the application of the operations, these distributions maximize and minimize the normalization ratio, respectively. 

\subsection{Uniforming and peaking operations}  \label{operations}
We construct \emph{uniforming} and \emph{peaking} operations on the distributions $\vec \lambda$ that act only on three selected $\lambda_j$, with the indices $1 \le j_1 < j_2 < j_3 \le S$. The operations will leave 
\eq
K_1&=&\lambda_{j_1} + \lambda_{j_2} + \lambda_{j_3} , \\ 
K_2&=&\lambda_{j_1}^2 + \lambda_{j_2}^2 + \lambda_{j_3}^2 ,
 \en
 invariant, and, consequently, also the total sum of the $\lambda_j$ and of the $\lambda_j^2$. The third power-sum, $M(3)$, however, will be changed by the operations. 
 
In analogy to an analysis of the birthday-problem with non-uniform birthday probabilities \cite{Munford1977}, we define the two operations, $\Gamma^{u}$ and $\Gamma^{p}$, on the probability distribution $\vec \lambda$: 
\eq 
\Gamma^{p/u}(\lambda_{j_1})&=&\frac{1}{3} \left(K_1 \pm \sqrt{6 K_2-2 K_1^2} \right) , \nonumber \\
\Gamma^{p/u}(\lambda_{j_2}) = \Gamma^{p/u}(\lambda_{j_3}) &=& \frac{1}{6} \left(2 K_1 \mp \sqrt{6 K_2-2 K_1^2} \right) \nonumber  ,\\
\Gamma^{p/u}(\lambda_{k \neq j_1, j_2, j_3})& = &\lambda_{k} , \label{ope1}
\en
where the upper (lower) sign in $\pm$ and $\mp$ refers to the peaking (uniforming) operation $\Gamma^p$ ($\Gamma^u$). For $K_1^2 <2 K_2$, we formally have $\Gamma^{u}(\lambda_{j_1})<0$, and we alternatively set
\eq 
\Gamma^{u}(\lambda_{j_1})&=&0 \nonumber , \\
\Gamma^{u}(\lambda_{j_2/j_3})&=&\frac{1}{2} \left( K_1 \pm \sqrt{2 K_2-K_1^2}\right) . \label{ope2}
\en
For convenience of notation, we set 
\eq 
\tilde \lambda^u_j=\Gamma^u(\lambda_j),  \
\tilde \lambda^p_j=\Gamma^p(\lambda_j) .
\en

\begin{figure}[ht]
\center
\includegraphics[width=\linewidth,angle=0]{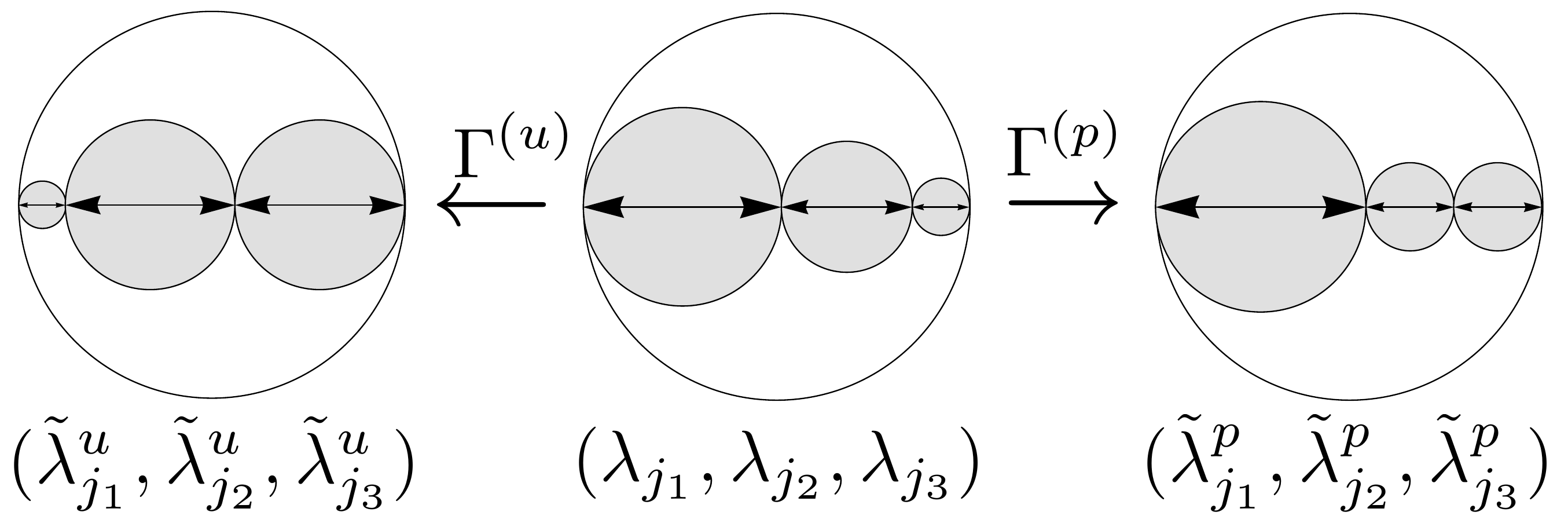} 
\caption{Action of uniforming and peaking operations $\Gamma^{u}$ and $\Gamma^{p}$. The tuple $(\lambda_{j_1}, \lambda_{j_2}, \lambda_{j_3})$ is leveled out by $\Gamma^{u}$, and made more peaked by $\Gamma^{p}$. We show only the coefficients $\lambda_j$ with indices $j_1, j_2, j_3$, all other coefficients $\lambda_k$ remain constant under the application of the operations, for the choice of indices $j_1, j_2, j_3$. The gray area represents the sum of the squared coefficients, it is the same for all three distributions.}
 \label{visuoperations}
\end{figure}

Colloquially speaking, $\Gamma^u$ \emph{levels out} the three coefficients $\lambda_{j_1}, \lambda_{j_2}, \lambda_{j_3}$ and thus makes the distribution more uniform, whereas $\Gamma^p$ makes the distribution more peaked. In both cases, the purity $P$ is kept constant. The operations push a distribution $\vec \lambda$ towards the uniform and peaked distribution, respectively, as illustrated in Fig.~\ref{visuoperations}. 

The uniform (peaked) distribution $\vec \lambda^{(u)}$ ($\vec \lambda^{(p)}$) is the only one that remains invariant under the application of $\Gamma^p$ ($\Gamma^u$), for all choices of $j_1, j_2, j_3$ (disregarding permutations of the indices), which can be seen by applying the operations on the distributions.

\subsection{Normalization ratio under operations} \label{normrationunder}
We now show that the uniforming (peaking) operation $\Gamma^{u(p)}$ reduces (increases) the normalization ratio, {\it i.e.}~we conjecture
\eq 
\frac{\chi_{N+1}(\Gamma^{u}(\vec \lambda) )}{\chi_N(\Gamma^{u}_{\phantom 0}(\vec \lambda) )} \le \frac{\chi_{N+1}(\vec \lambda)}{\chi_N(\vec \lambda)} \le \frac{\chi_{N+1}(\Gamma^{p}(\vec \lambda) )}{\chi_N(\Gamma^{p}_{\phantom o}(\vec \lambda) )} , \label{operationinequality}
\en
where we made the dependence of $\chi_N$ on the distributions explicit.

Without restrictions of generality and for convenience of notation, we set $j_1=1, j_2=2, j_3=3$, {\it i.e.}~we let the operations act on the first three Schmidt coefficients.

We  define 
\eq 
\tilde \chi_N&=&\chi_N(\lambda_4, \dots, \lambda_S),  \\
\Lambda& =& \lambda_1  \lambda_2  \lambda_3 
\en
{\it i.e.}~formally $\tilde \chi_N$ is a normalization factor, but for an unnormalized distribution $\{ \lambda_4, \dots, \lambda_S \}$. With these definitions, we rewrite $\chi_N$ as 
\eq \chi_N &=& \Lambda \cdot  \tilde \chi_{N-3} +(\lambda_1 \lambda_2 +\lambda_3 \lambda_2+\lambda_1 \lambda_3)  \tilde \chi_{N-2}  \nonumber \\
&& 
 +(\lambda_1+\lambda_2+\lambda_3) \tilde \chi_{N-1}+ \tilde \chi_{N} . \label{rewritechn} \en
The terms \eq 
\lambda_1 \lambda_2 +\lambda_3 \lambda_2 +\lambda_1 \lambda_3 &=& \frac 1 2 \left( K_1^2 - K_2 \right) , \\ 
\lambda_1+\lambda_2+\lambda_3&=& K_1 ,   \en 
and $\tilde \chi_k$ with $k \in \{N-3 , \dots,  N \}$ do not change upon  application of $\Gamma^{u/p}$, {\it i.e.}~only the product $\Lambda$ is affected by the operations.

{\it Conjecture:} It holds 
\eq 
\tilde \lambda^u_1 \tilde \lambda^u_2 \tilde \lambda^u_3  \le \Lambda \le \tilde  \lambda^p_1 \tilde \lambda^p_2 \tilde \lambda^p_3 , \label{lambdaprohier1}
\en
where $\tilde \lambda_j^{u/p}$ is the result of the operation $\Gamma^{u/p}$ on $\lambda_j$.

{\it Proof:} We re-write the products in terms of $K_1, K_2$ and $\lambda_1$
\begin{figure}[h]
\center
\includegraphics[width=.90\linewidth,angle=0]{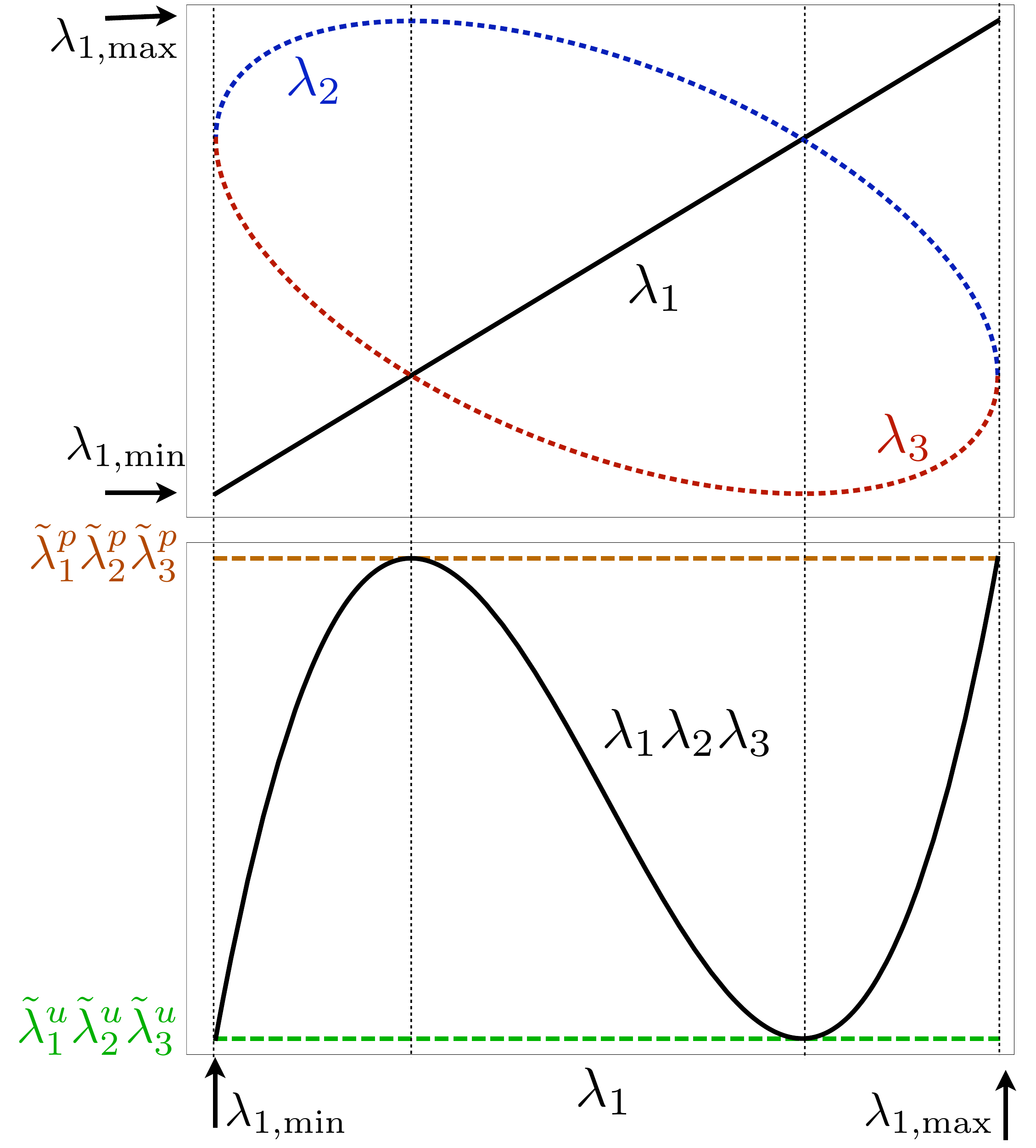} 
\caption{(color online) Upper panel: Possible values of $\lambda_1$, $\lambda_2$ and $\lambda_3$, given $K_1$ and $K_2$, as a function of $\lambda_1$, in arbitrary units. Lower panel: Behavior of $\lambda_1 \lambda_2 \lambda_3$ as a function of $\lambda_1$, and lower and upper bounds. The lower (upper) bounds are attained if and only if the configuration fulfills $\lambda_{j_1} > \lambda_{j_2} = \lambda_{j_3}$ ($\lambda_{j_1} < \lambda_{j_2} = \lambda_{j_3}$), where $\{ j_1, j_2, j_3\}$ is a permutation of $\{1,2,3\}$. These points are marked by thin dotted vertical lines. } 
 \label{illulambdaprod}
\end{figure}

\eq 
\nonumber
\tilde \lambda^u_1 \tilde \lambda^u_2 \tilde \lambda^u_3 &=&  \left\{ \begin{tabular}{ll}
 $\frac{1}{108} \left(K_1-\sqrt{6 K_2-2 K_1^2}\right) $  & for $ K_1^2 > 2 K_2$ \\ 
$\times \left(2 K_1+\sqrt{6 K_2-2 K_1^2}\right)^2$ & \vspace{0.21cm} \\ 
0 & for $ K_1^2 \le 2 K_2$  \end{tabular} \right. \\
\Lambda&=& \frac{1}{2} \lambda_1 \left(2 \lambda_1^2-2 \lambda_1 K_1+K_1^2-K_2 \right)  \\
\tilde \lambda^p_1 \tilde \lambda^p_2 \tilde \lambda^p_3 &=& \frac{1}{108}  \left( K_1+\sqrt{6 K_2-2 K_1^2}\right) \nonumber \\ &&  \times \left(2 K_1-\sqrt{6 K_2-2 K_1^2}\right)^2 \nonumber
\en

For given $K_1$ and $K_2$,  we can find the possible  $\lambda_{2/3}$, leaving $\lambda_1$ as a free parameter,
\eq
\lambda_{2/3} = \frac 1 2 \left( K_1 - \lambda_1 \pm \sqrt{2 \lambda_1 K_1 - K_1^2 - 3 \lambda_1^2 + 2 K_2} \right) \nonumber \en
The requirement $\lambda_{2/3} \ge0$ gives 
\eq
\lambda_{1,{\text{max}/\text{min}}} &=&\frac 1 3 \left( K_1 \pm \sqrt 2 \sqrt{3 K_2 - K_1^2} \right) , \\
 \lambda_{1,{\text{min}}} &\le& \lambda_1 \le \lambda_{1,{\text{max}}}
\en
Given $K_1, K_2$, all possible values of $\lambda_1$ then fulfill Eq.~(\ref{lambdaprohier1}).  \qed

The inequalities in (\ref{lambdaprohier1}) are saturated for $\tilde \lambda^{u/p}_j=\lambda_j$ (modulo permutation of the indices). Possible values of $\lambda_{j}$ and the behavior of the product  $\lambda_1 \lambda_2 \lambda_3=\Lambda$ are shown in Fig.~\ref{illulambdaprod}. With (\ref{rewritechn}), it immediately follows that 
\eq 
\chi^{(u)}_{N}  \le {\chi^{\phantom u}_{N}} \le {\chi^{(p)}_{N}} .
\en
Using (\ref{lambdaprohier1}), we can set
\eq 
\lambda^p_1 \lambda^p_2 \lambda^p_3 =: \Lambda (1+\epsilon),  \nonumber  \\
\lambda^u_1 \lambda^u_2 \lambda^u_3 =: \Lambda (1-\delta), \nonumber  
\en
with $\epsilon, \delta \ge 0$, and  
\eq 
B_N&=&(\lambda_1 \lambda_2 +\lambda_3 \lambda_2+\lambda_1 \lambda_3)  \tilde \chi_{N-2}  \nonumber \\
&& +(\lambda_1+\lambda_2+\lambda_3) \tilde \chi_{N-1}+ \tilde \chi_{N}   \nonumber  .
\en
For any distribution $\vec \lambda$ (which does not need to fulfill $\sum_j \lambda_j=1$), the Newton-Maclaurin inequality holds \cite{Macmahon1915,Chudzicki2010}, which reads 
\eq 
\frac{\tilde \chi_{N+1}}{\tilde \chi_N} \le \frac{\tilde \chi_{N}}{\tilde \chi_{N-1}} .
\en
We thus have
\eq 
\tilde \chi_{N-3} \ge \tilde \chi_{N-2} \label{chiine}, \ B_{N} \ge B_{N+1} .\label{Biine}
\en
Our original conjecture (\ref{operationinequality}) is equivalent to 
\eq 
\frac{\Lambda(1-\delta) \tilde \chi_{N-2} + B_{N+1} }{\Lambda(1-\delta) \tilde \chi_{N-3} + B_{N} } \le  
\frac{\Lambda \tilde \chi_{N-2} + B_{N+1} }{\Lambda  \tilde \chi_{N-3} + B_{N} } \nonumber \\
\le 
\frac{\Lambda(1+ \epsilon ) \tilde \chi_{N-2} + B_{N+1} }{\Lambda(1+ \epsilon) \tilde \chi_{N-3} + B_{N} }  ,
\en
and follows from (\ref{chiine}). 

Consequently, $\vec \lambda^{(p)}$ ($\vec \lambda^{(u)}$) maximizes (minimizes) the normalization ratio $\chi_{N+1}/\chi_{N}$ for a given $P$ and $S$, which proves $(ii)$ and $(iii)$. The inequality $(iv)$ then follows by taking the indicated limit, $S\rightarrow \infty$.

\end{document}